\documentclass[onecolumn,showpacs,preprintnumbers,showkeys,amsmath,amssymb]{revtex4}

\usepackage{graphicx}
\usepackage{dcolumn}
\usepackage{bm}

\begin{document}

\title{The CrMES scheme as an alternative to Importance Sampling: The tail regime of the order-parameter distribution}

\author{Anastasios Malakis}
\altaffiliation[]{Corresponding author: amalakis@cc.uoa.gr}
\author{Nikolaos G. Fytas}
\affiliation{Department of Physics, Section of Solid State
Physics, University of Athens, Panepistimiopolis, GR 15784
Zografos, Athens, Greece}

\date{\today}

\begin{abstract}
We review the recently developed critical minimum energy-subspace
(CrMES) technique. This scheme produces an immense optimization of
popular algorithms, such as the Wang-Landau (WL) and broad
histogram methods, by predicting the essential part of the energy
space necessary for the estimation of the critical behavior and
provides a new route of critical exponent estimation. A powerful
and efficient CrMES entropic sampling scheme is proposed as an
alternative to the traditional importance sampling methods.
Utilizing the WL random walk process in the dominant energy
subspace (CrMES-WL sampling) and using the WL approximation of the
density of states and appropriate microcanonical estimators we
determine the magnetic properties of the 2D Ising model. Updating
$(E,M)$ histograms during the high level WL-iterations, we provide
a comprehensive alternative scheme to the Metropolis algorithm and
by applying this procedure we present a convincing analysis for
the far tail regime of the order-parameter probability
distribution.
\end{abstract}

 \maketitle

\section{Introduction}
\label{section1}

Traditional Monte Carlo sampling methods have increased
dramatically our understanding of the behavior of the standard
classical statistical mechanics systems. The Metropolis method and
its variants were, for many years, the main tools in condensed
matter physics, particularly for the study of critical
phenomena~\cite{metropolis53,binder77,landau00}. However, in many
cases, such as for example in complex systems with effective
complicated potentials, the Metropolis method and its variants are
more or less inadequate methods~\cite{landau00,newman99}.
Noteworthy, the traditional importance sampling methods are not
very efficient in recording of the very small probabilities in the
tails of the critical order-parameter distribution~\cite{smith95}.
This is the reason why many different simulation
approaches~\cite{smith95,hilfer95,stauffer98,tsypin00} in recent
times have failed in establishing the true tail behavior of this
distribution.

In Sec.~\ref{section2}, we briefly review a new simple technique,
the CrMES method, which yields an immense speed up of all popular
algorithms, which are used to sample the density of energy states
(DOS) of a statistical system in the last decade. Provided that,
the temperature range of interest is only the range around the
critical temperature that determines all finite-size anomalies,
our technique can be combined with any of the DOS
methods~\cite{landau00,newman99,lee93,berg92,lima00,oliveira00,kastner00,wang02,wang01,landau04}
to speed up the simulations and establish a different methodology
for the determination of the critical exponents by studying the
finite-size scaling of the extensions of the dominant subspaces.
In the same section we describe a CrMES-WL entropic sampling
method and all magnetic properties are obtained by using the
high-levels of the WL random walk process to determine appropriate
microcanonical estimators. The proposed method is combined with
the N-fold way~\cite{bortz75,malakis04b,schulz01} in order to
improve efficiency and statistical reliability. Finally, in
Sec.~\ref{section3} we study the order-parameter distribution of
the 2D Ising model and we clarify its asymptotic behavior in the
far tail regime. Our conclusions are summarized in
Sec.~\ref{section4}.

\section{The CrMES Wang-Landau entropic sampling scheme}
\label{section2}

According to the CrMES
method~\cite{malakis04,martinos05,malakis05} the total energy
range $(E_{min},E_{max})$ is restricted using the definitions:
\begin{equation}
\label{eq:1} (\widetilde{E}_{-},\widetilde{E}_{+}),\;\;\;\;
\widetilde{E}_{\pm}=\widetilde{E}\pm \Delta^{\pm},\;\;\;\;\;
\Delta^{\pm}\geq0
\end{equation}
with respect to the value $\widetilde{E}$ producing the maximum
term in the partition function of the statistical model, for
instance the Ising model, at some temperature of interest. Thus,
the specific heat peaks are approximated by:
\begin{equation}
\label{eq:2}
C_{L}^{\ast}(\widetilde{E}_{-},\widetilde{E}_{+})\equiv
C_{L}^{\ast}(\Delta^{\pm})=N^{-1}T^{-2}\left\{\widetilde{Z}^{-1}
\sum_{\widetilde{E}_{-}}^{\widetilde{E}_{+}}E^{2}\exp{[\widetilde{\Phi}(E)]}-
\left(\widetilde{Z}^{-1}\sum_{\widetilde{E}_{-}}^{\widetilde{E}_{+}}E
\exp{[\widetilde{\Phi}(E)]}\right)^{2}\right\}
\end{equation}

\begin{equation}
\label{eq:3} \widetilde{\Phi}(E)=[S(E)-\beta
E]-\left[S(\widetilde{E})-\beta
\widetilde{E}\right],\;\;\widetilde{Z}=\sum_{\widetilde{E}_{-}}^{\widetilde{E}_{+}}\exp{[\widetilde{\Phi}(E)]}
\end{equation}

\begin{figure}[htbp]
\includegraphics{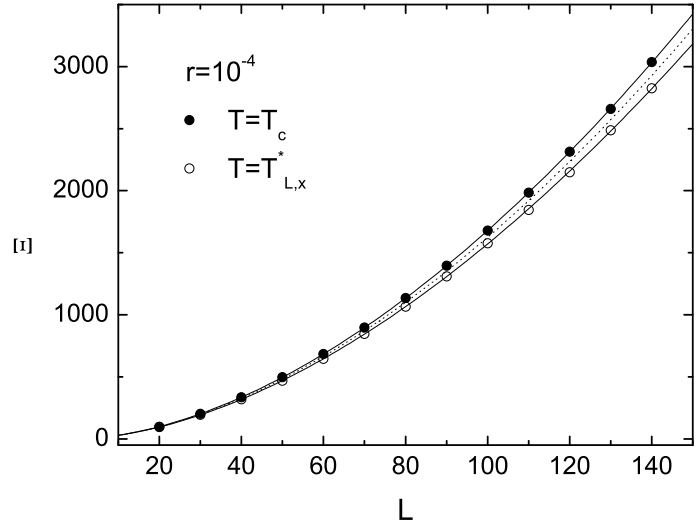}
\caption{\label{fig1}Behavior of CrMMS for $r=10^{-4}$. The solid
lines correspond to power law fits (a) for $T=T_{c}$: $\Xi\simeq
0.513(15)\cdot L^{1.758(7)}$ and (b) for $T=T_{L,\chi}^{\ast}$:
$\Xi\simeq 0.517(45)\cdot L^{1.741(18)}$. The dotted line
represents the law $\Xi\equiv 0.5138\cdot L^{1.75}$, which is very
close to the fit $\Xi\simeq 0.5145(90)\cdot L^{1.7497(50)}$
corresponding to the average
$(\Xi_{T_{c}}+\Xi_{T_{L,\chi}^{\ast}})/2$.}
\end{figure}
To implement the above restriction we request a specified accuracy
by imposing the condition:
\begin{equation}
\label{eq:4}
\left|\frac{C_{L}^{\ast}(\Delta_{\pm})}{C_{L}^{\ast}}-1\right|\leq
r
\end{equation}
where $r$ measures the relative error and will be set equal to a
small number ($r=10^{-4}$ and/or $r=10^{-6}$), and $C_{L}^{\ast}$
is the value of the maximum of the specific heat obtained by using
the total energy range. With the help of a convenient
definition~\cite{malakis04,malakis05}, we can specify the minimum
energy subspace satisfying the above condition. In fact, it has
been numerically verified that the finite-size extensions (denoted
by $\Delta\widetilde{E}\equiv
\min(\widetilde{E}_{+}-\widetilde{E}_{-})$) close to a critical
point obey the scaling law~\cite{malakis04}:
\begin{equation}
\label{eq:5} \Psi_{C_{L}^{\ast}}\equiv
\frac{(\Delta\widetilde{E})^{2}_{C_{L}^{\ast}}}{L^{d}}\approx
L^{\frac{\alpha}{\nu}}
\end{equation}
To locate the CrMES we may follow the method described
in~\cite{malakis04}, or an even simpler
restriction~\cite{malakis05} based on the energy probability
density $(f_{T_{L}^{\ast},C}(E)\propto \widetilde{\Phi}(E))$.
Using this later approach, we may define the end-points
$(\widetilde{E}_{\pm})$ of the subspaces by simply comparing the
corresponding probability densities with the maximum at the energy
$\widetilde{E}$:
\begin{equation}
\label{eq:6} \widetilde{E}_{\pm}:\;\;\;
\exp{\{\widetilde{\Phi}(\widetilde{E}_{\pm})\}}\leq r
\end{equation}
Alternative definitions for the CrMES have been described and
tested in Ref.~\cite{malakis05} using the 2D Ising
model~\cite{malakis04,malakis05}, the 3D Ising
model~\cite{malakis04,malakis05} and the Baxter-Wu
model~\cite{martinos05,malakis05}.

\begin{figure}[htbp]
\includegraphics{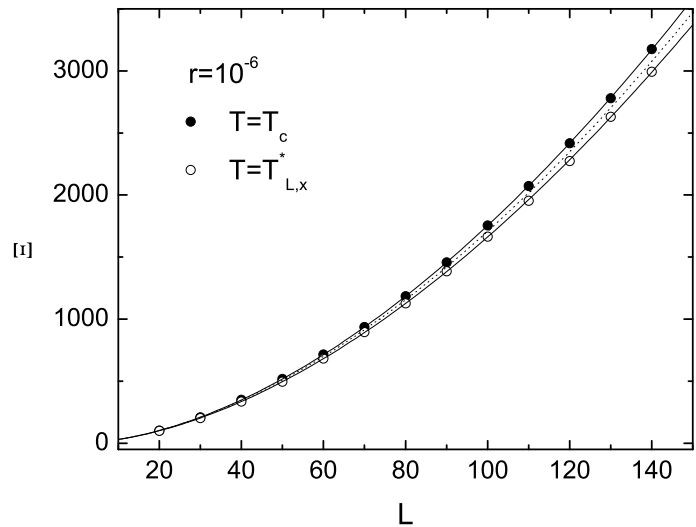}
\caption{\label{fig2}Behavior of CrMMS for $r=10^{-6}$. The solid
lines correspond to power law fits (a) for $T=T_{c}$: $\Xi\simeq
0.528(17)\cdot L^{1.761(7)}$ and (b) for $T=T_{L,\chi}^{\ast}$:
$\Xi\simeq 0.536(45)\cdot L^{1.745(15)}$. The dotted line
represents the law $\Xi\equiv 0.54\cdot L^{1.75}$, which is very
close to the fit $\Xi\simeq 0.532(15)\cdot L^{1.753(6)}$
corresponding to the average
$(\Xi_{T_{c}}+\Xi_{T_{L,\chi}^{\ast}})/2$.}
\end{figure}
Let us now discuss the idea of producing accurate estimates for
finite-size magnetic anomalies by using a simple method based on a
WL random walk in an appropriately restricted energy subspace
$(E_{1},E_{2})$. Implementing this scheme we, at the same time,
accumulate data for the two-parameter $(E,M)$ histogram. A
multi-range algorithm~\cite{wang01} is implemented to obtain the
DOS and the $(E,M)$ histograms in $(E_{1},E_{2})$. The WL
modification factor $(f_{j})$ is reduced at the jth iteration
according to: $f_{1}=e,\; f_{j}\rightarrow f_{j-1}^{1/2},\;
j=2,...,J_{fin}$. The approximation of the DOS, in the last WL
iteration, $G_{WL}(E)$, and the high-level $(j\gg1)$ WL $(E,M)$
histograms, $H_{WL}(E,M)$, are used to estimate the magnetic
properties in a temperature range, which is covered, by the
restricted energy subspace $(E_{1},E_{2})$ as:
\begin{equation}
\label{eq:7} \langle M^{n}\rangle=\frac{\sum_{E}\langle
M^{n}\rangle_{E}G(E)e^{-\beta E}}{\sum_{E}G(E)e^{-\beta E}}\cong
\frac{\sum_{E\in(E_{1},E_{2})}\langle
M^{n}\rangle_{E,WL}G_{WL}(E)e^{-\beta
E}}{\sum_{E\in(E_{1},E_{2})}G_{WL}(E)e^{-\beta E}}
\end{equation}
The microcanonical averages $\langle M^{n}\rangle_{E}$ are
obtained from the $H_{WL}(E,M)$ histograms as:
\begin{equation}
\label{eq:8}\langle M^{n}\rangle_{E}\cong\langle
M^{n}\rangle_{E,WL}\equiv
\sum_{M}M^{n}\frac{H_{WL}(E,M)}{H_{WL}(E)},\;\;H_{WL}(E)=\sum_{M}H_{WL}(E,M)
\end{equation}
and the summation in the magnetization $M$ runs over all values
generated in the restricted energy subspace $(E_{1},E_{2})$.

The accuracy of the magnetic properties obtained from the above
averaging process depends on many factors. However, since the
detailed balance condition depends on the control parameter
$(f_{j})$, we classify our recipes utilizing the $j$-range used
for updating the $(E,M)$ histogram during the WL process. The
high-level recipes WL$(J_{init},J_{fin})$ and their N-fold
versions WL(N-fold:$J_{N-fold},J_{fin})$ give excellent estimates,
as shown in detail in Ref.~\cite{malakis05}. The extensions of the
CrMES defined with the help of the thermal finite-size anomalies
(specific heat maximum and energy cumulant minimum), and also with
the help of magnetic finite-size anomaly (susceptibility maximum)
obey a clear logarithmic dependence on the lattice
size~\cite{malakis05}, as should be expected for the 2D Ising
model.

Consider now the probability density of the order-parameter at
some temperature of interest $T$:
\begin{equation}
\label{eq:9}
P_{T}(M)\cong\frac{\sum_{E\in(E_{1},E_{2})}\frac{H_{WL}(E,M)}{H_{WL}(E)}G_{WL}(E)e^{-\beta
E}}{\sum_{E\in(E_{1},E_{2})}G_{WL}(E)e^{-\beta E}}
\end{equation}
If $\widetilde{M}$ is the value that maximizes (\ref{eq:9}), we
locate the critical minimum magnetic subspaces (CrMMS) by:
\begin{equation}
\label{eq:10} \widetilde{M}_{\pm}:\;\;\;
\frac{P_{T_{L,\chi}^{\ast}}(\widetilde{M}_{\pm})}{P_{T_{L,\chi}^{\ast}}(\widetilde{M})}\leq
r
\end{equation}
and  we should expect the following scaling law to hold:
\begin{equation}
\label{eq:11} \Xi_{T_{L,\chi}^{\ast}}\equiv
\Xi_{P_{T_{L,\chi}^{\ast}}(M)}\equiv \frac{(\Delta
\widetilde{M})^{2}_{T_{L,\chi}^{\ast}}}{L^{d}}\approx
L^{\gamma/\nu}
\end{equation}

\begin{figure}[htbp]
\includegraphics{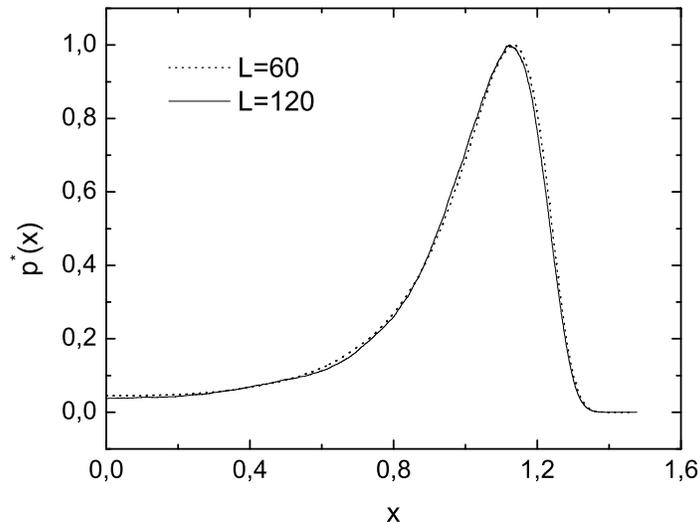}
\caption{\label{fig3}Illustration of the universal scaling
function $p^{\ast}(x)$ for $L=60$ and $L=120$.}
\end{figure}
In Figs.~\ref{fig1},\ref{fig2} we present the behavior of the
extensions of the CrMMS, obtained using the magnetic space
restriction (\ref{eq:10}) for two values of $r$, namely
$r=10^{-4}$ and $r=10^{-6}$. These extensions were determined at
the exact critical temperature $T_{c}$ and at the pseudocritical
temperatures $T_{L,\chi}^{\ast}$ of the susceptibility, as
indicated in the figures. The power laws and the resulting
exponents are given in the corresponding figure captions. It is
clear that the law (\ref{eq:11}) is very well satisfied and the
estimation of $\gamma/\nu$ via this route is quite good, giving
$\gamma/\nu=1.7497(50)$ for $r=10^{-4}$ and $\gamma/\nu=1.753(6)$
for $r=10^{-6}$, respectively. It is of interest to point out
that, an attempt to estimate $\gamma/\nu$ using this route and the
Metropolis algorithm will most likely yield a marked
underestimation, which will be more pronounced for the case
$r=10^{-6}$. This effect, has been discussed in
Ref.~\cite{malakis05} and is a result of the very slow
equilibration process of this algorithm in the far tail regime of
the order-parameter distribution. Next, we consider the tail
regime of the universal order-parameter distribution, we analyze
its asymptotic behavior and finally examine the possibility of
extracting estimates of the exponent $\delta$ from this behavior.

\section{Far tail regime of the order-parameter distribution}
\label{section3}

Following Smith and Bruce~\cite{smith95} we define the universal
scaling form of the order-parameter density by:
\begin{equation}
\label{eq:12} p^{\ast}(x)dx\simeq p(m)dm,\;\; x=m/\sqrt{\langle
m^{2} \rangle},\;\; m=M/N
\end{equation}

Fig.~\ref{fig3} shows $p^{\ast}(x)$ for $L=60$ and $L=120$
obtained via the WL(N-fold:12-24) scheme. The tail regime of this
universal distribution has been a matter of increasing interest in
the last decade~\cite{smith95,hilfer95,hilfer94,bruce95,hilfer03}.
The main theme is the verification of the following conjecture for
the large-$x$ behavior of $p^{\ast}(x)$~\cite{smith95,hilfer95}:
\begin{equation}
\label{eq:13} p^{\ast}(x)\simeq
p_{\infty}x^{\psi}\exp(-a_{\infty}x^{\delta+1})
\end{equation}
with:
\begin{equation}
\label{eq:14} \psi=\frac{\delta-1}{2}
\end{equation}
and $p_{\infty}$, $a_{\infty}$ universal constants. The above
hypothesis, its origin and its significance are discussed in some
detail in Refs.~\cite{smith95,hilfer95}.
\begin{figure}[htbp]
\includegraphics{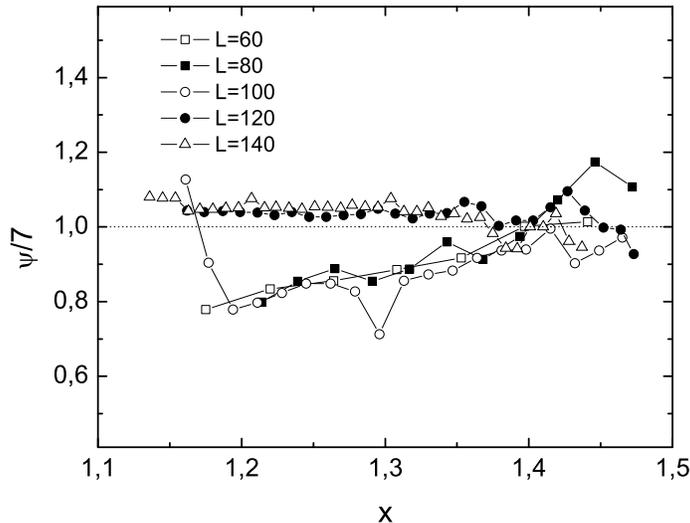}
\caption{\label{fig4}Behavior of estimates of the exponent $\psi$
for lattice sizes $L=60-140$. The points correspond to the ratio
of the best fit value of the exponent to the expected value $7$
(to be compared with Fig. 7(b) of Ref.~\cite{smith95}).}
\end{figure}

The studies of~\cite{smith95,hilfer95} have provided some evidence
for this conjecture and in particular for the prefactor and the
relation of the exponent $\psi$ to the critical exponent $\delta$.
For the 2D Ising model the exponent should have the value
$\psi=7$, if, of course, the prefactor hypothesis is valid. Smith
and Bruce~\cite{smith95} provided numerical support for this
value, but their study was not completely conclusive since it was
carried out only for relatively small lattices ($L=32$ and $L=64$)
and the $x$-window in which the value $\psi=7$ was observed was
actually quite narrow. We now present results for several lattice
sizes ($L=80,100,120$, and $L=140$) reinforcing this conjecture in
a very wide $x$-window. Following Smith and Bruce~\cite{smith95}
we fix the exponent $\delta$ in the exponential factor of
Eq.~(\ref{eq:13}) and fit our results $(x>1)$ in $x$-windows, each
one corresponding to $50$ different magnetization values, sampled
during the WL(N-fold:12-24) process. Fig.~\ref{fig4} shows a very
clear signature of the prefactor law (\ref{eq:14}) which upholds
in a large $x$-window only for the large lattices ($L=120$ and
$L=140$). On the other hand, for smaller lattice sizes
($L=60-100$) the picture is similar to that presented in
Ref.~\cite{smith95} and the expected value is obtained only in a
small $x$-window. Finally, let us treat both exponents in the
exponential and in the prefactor as free parameters, assuming
however the validity of Eq.~(\ref{eq:14}). In this case we examine
whether the far tail regime could be a possible route for an
independent estimation of the exponent $\delta$.
\begin{figure}[htbp]
\includegraphics{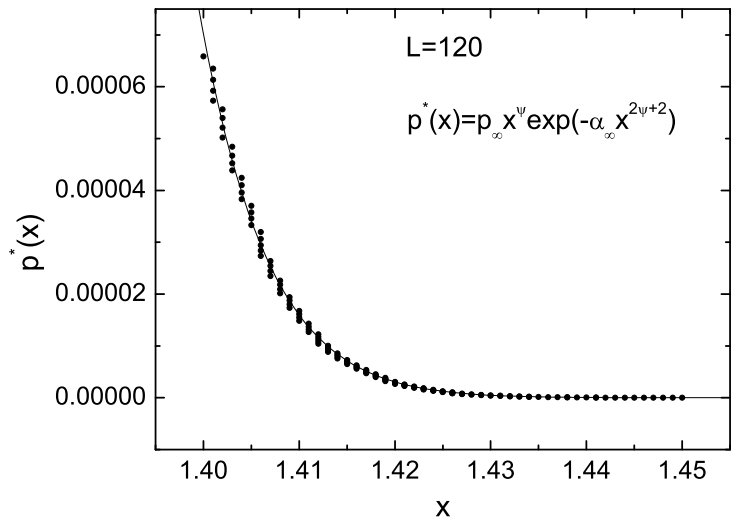}
\caption{\label{fig5}Independent estimation of the exponent
$\delta$ via fitting the universal scaling function $p^{\ast}(x)$
in a specific $x$-window. The fitted line reads as
$p^{\ast}(x)=p_{\infty}x^{\psi}\exp(-a_{\infty}x^{2\psi+2})$,
yielding $p_{\infty}=2.189(2.223)$, $\psi=6.995(1.750)$ and
$a_{\infty}=0.058(64)$, respectively.}
\end{figure}
Fig.~\ref{fig5} presents such an attempt for the case $L=120$. The
value obtained for $\psi$ (and therefore for $\delta(=2\psi+1)$)
is accurate to the third decimal place $(\psi=6.995(1.750))$,
despite the fact of the existing large error of the fit. In fact
the shown $x$-window $(x=1.4-1.45)$ produces good estimates for
all lattice sizes $L=60-140$ as should be expected by thoroughly
inspecting Fig.~\ref{fig4}.

\section{Concluding remarks}
\label{section4}

In the present work we described the basic ideas and formalism
behind the recently developed by the
authors~\cite{malakis04,martinos05,malakis05} critical minimum
energy(magnetization) subspace CrME(M)S technique, appearing as an
alternative computational method for the estimation of critical
behavior of statistical systems. We presume that this new
promising route will increase our comprehension of the development
of the critical behavior as the lattice size increases and will
further facilitate the estimation of critical exponents via
finite-size scaling. It is hoped that the presented CrMES
Wang-Landau entropic sampling will provide an efficient and
reliable scheme for the study of complex systems, such as the
random-field Ising model and systems with competing interactions,
where the critical behavior is still controversial. Furthermore,
the efficiency of our approach enabled us to present here reliable
data for large lattices and to clarify the asymptotic tail
behavior of the universal critical distribution of the
order-parameter of the 2D Ising model. This should be contrasted
to the fact that all previous studies of the far tail regime of
this distribution~\cite{smith95,hilfer03} have considered rather
small lattices.

\begin{acknowledgments}
{This research was supported by EPEAEK/PYTHAGORAS under Grant No.
$70/3/7357$.}
\end{acknowledgments}

{}

\end{document}